# Further Decimating the Inductive Programming Search Space with Instruction Digrams

Edward Mc Daid[a] and Sarah Mc Daid[b]
*Zoea Ltd., 20-22 Wenlock Road, London, N1 7GU, UK*
*{edward.mcdaid, sarah.mcdaid}@zoea.co.uk*

Abstract: Overlapping instruction subsets derived from human originated code have previously been shown to dramatically shrink the inductive programming search space, often by many orders of magnitude. Here we extend the instruction subset approach to consider direct instruction-instruction applications (or instruction digrams) as an additional search heuristic for inductive programming. In this study we analyse the frequency distribution of instruction digrams in a large sample of open source code. This indicates that the instruction digram distribution is highly skewed with over 93% of possible instruction digrams not represnted in the code sample. We demonstrate that instruction digrams can be used to constrain instruction selection during search, further reducing size of the the search space, in some cases by several orders of magnitude. This significantly increases the size of programs that can be generated using search based inductive programming techniques. We discuss the results and provide some suggestions for further work.

## 1 INTRODUCTION

The production of software remains a complex, specialist, labour intensive and expensive activity. This is despite many (mostly incremental) advances in programming language design and software development methods [1]. In this context the idea of using of AI to simplify or even automate coding is appealing and much work has been directed to this end [2,3].

Recently, there has been significant interest and progress in the application of deep learning and large language models to the production of code [4]. While promising, these approaches are not directly relevant to the work described in this paper and will not be discussed here in further detail. Instead, our focus is inductive programming - a field that has been active over a long period [5].

The goal of inductive programming is to transform a specification - such as a set of test cases - directly into software. Within this domain a variety of approaches have been developed [6] but fundamentally these are all limited by the size of the search space [7].

Most high-level programming languages include around 200 instructions, comprising operators as well as core and standard library functions. In simplistic terms this dictates the branching factor of the inductive programming search space - which grows exponentially with increasing target program size.

Aside from trivial cases it is not possible to determine the output of a given source code program without executing it. As a result, all inductive programming approaches rely on some form of generate and test. The huge size of this search space has meant that until recently inductive programming was limited to the production of small programs [7].

Zoea is an inductive programming system that has been developed over the last few years [8]. To some extent Zoea sidesteps the inductive programming search problem by avoiding search where possible and through the paradigm of composable inductive programming (CIP) [9]. CIP is an iterative and incremental process that involves the composition of small program units, which are generated using inductive programming. This is facilitated by a visual programming language called Zoea Visual [10]. In principle CIP allows inductive programming to be used to incrementally produce software of any size [11]. However, the size of individual program units is still limited in some cases by search.

---

[a] https://orcid.org/0000-0001-8684-0868
[b] https://orcid.org/0000-0001-7643-6722

Zoea employs a blackboard architecture that supports clustered deployment [12]. Zoea knowledge sources also support partitioning of the search space using subsets of the instruction set. Recently, a study was conducted to produce new instruction subsets capable of supporting hundreds of cores [13]. This involved the creation of thousands of small overlapping instruction subsets. These subsets were derived from a large sample of open source code. Subsets are created by clustering the program unit instruction subsets from the code sample. It was found that the derived instruction subsets generalised quickly to cover unseen code. They also effectively reduce the overall size of the search space - often by many orders of magnitude.

Instruction subsets shrink the search space due to the skewed frequency distribution of individual instructions and the even more skewed distribution of instruction co-occurrence within program units. Human developers do not use all instructions or combinations of instructions with equal probability and the skewed patterns of instruction co-occurrence in human code persist in the derived subsets. As such, instruction subsets represent a form of low-level tacit programming knowledge [14]. For inductive programming they also represent a useful and general heuristic [15].

The previous study noted the highly skewed distribution of instruction pair co-occurrence within program units. This raises the possibility that the frequency distribution of direct instruction application might represent another useful heuristic.

In Zoea, candidate solution programs are assembled by combining instructions to form a functional program that matches the test cases. Factors such as the data type compatibility of instruction inputs and outputs already constrain which instructions can be applied in many cases. Also, we know from our work with instruction subsets that many pairs of individual instructions are never used together by human developers within the same program unit - or are very rarely used. If we understood how human developers combine individual instructions to form programs then we could use this information to inform our choice of which instruction to apply next for any partially constructed solution candidate.

An instruction digram is simply an ordered pair of instruction identifiers that correspond to the direct application of one instruction to the return value of another instruction. The set of all possible instruction digrams can easily be enumerated from any set of instructions. For example, if the set of instructions is { a … z } then the set of all possible instruction digrams is { [a,a], [a,b] … [z,y], [z,z] }. Each possible instruction digram has a frequency distribution across all of the programs in any code sample – which may be zero.

If human developers use only a subset of all possible instruction digrams, then we can easily identify that subset and use it as an additional search constraint. As with instruction subsets, instruction digrams can be considered a form of heuristic. Instruction digrams could be used as a heuristic in their own right but the intention is to combine their use with that of instruction subsets to yield a larger reduction in search space size.

The key question addressed here is: does the use of instruction digrams reduce the branching factor and thus the size of the inductive programming search space, and if so by how much?

## 2 APPROACH

This study reused the same data set that was used for the creation of instruction subsets [13]. This consists of a snapshot of the largest 1000 repositories on GitHub [16] that was taken on 13 May 2022. Only python [17] code is used and the data set includes approximately 14.75 million lines of code in that language. Python is a popular language with an instruction set that is typical of many similar languages. It is also relatively easy to parse.

Identifying instruction digrams is more involved than the production of instruction subsets as it requires parsing rather than tokenisation. The python AST (abstract syntax tree) module [18] would be a natural choice for this task however it is not backwards compatible and as such is unable to process different python versions at the same time. The code sample contains python source code in a variety of versions.

Instead, a simple expression parser for python was constructed. Pre-processing with regular expressions filtered out comments and any code that did not contain multiple instructions. Also, literal strings were replaced with variable identifiers in order to prevent their contents from interfering with the parsing process. This resulted in a series of code fragments containing instructions that were then parsed individually.

A separate parse tree was produced for each code fragment that corresponded to the application of two or more instructions. The parse tree includes a node for each instruction (operator or function) together with links for each input or argument. Such a parse tree can include any number of instruction digrams - corresponding to any two directly connected nodes. Once constructed, each node in the parse tree was visited in turn, to identify all linked pairs of instruction nodes. The instruction digram patterns that were identified include the following examples:

- **function( function() )**
- **function().function()**
- <**expression**> <**operator**> **function()**
- **function()** <**operator**> <**expression**>
- **function(** <expression> <**operator**> <expression> **)**
- <expression> <**operator**> <expression> <**operator**> <expression>

An ordered pair of instruction identifiers [F1, F2] uniquely identifies each digram, where instruction F2 uses the output value of instruction F1 as an input. In temporal terms, F1 is executed before F2. Any occurrence of [F2, F1] is treated as a different digram.

For each source code file, the number of occurrences of each distinct digram was recorded. Once all files had been processed these numbers were totalled to produce counts for each digram across all files in a given repo. A similar roll up produced totals for each digram across all repos in the code sample.

A high level summary of the process for identifying instruction digrams is as follows:

1. The zip file for each repo was unpacked into a temporary folder;
2. All non-python code and data was deleted;
3. Each python source file was pre-processed and split into expressions;
4. Each expression was parsed to produce an AST;
5. Each AST was walked to identify every digram;
6. Counts for each unique digram for each file were produced;
7. Total instruction digram counts across all repos were produced.

A skewed digram frequency distribution would be required for digrams to represent a useful search heuristic. Ideally, this would include a significant number of digrams that are rarely or never used.

We used the set of digrams identified in the code sample to produce new search space size estimates and compared these with the results of the previous study. The objective here was to determine whether we obtained another significant reduction in the size of the search space using instruction digrams.

Estimation of the search space size was accomplished by using the digrams to determine which members of the instruction subset could be applied at each level in a search tree, up to some specified maximum depth. In doing this we only considered digrams where both instructions were present in each instruction subset.

At the first level of the search tree it was assumed that all subset instructions could be applied. At each subsequent level an instruction F2 could only be applied if a digram [F1, F2] existed, such that instruction F1 was present on a previous level and instruction F2 was a member of the current instruction subset. The size of the search space was measured as the total number of instructions present on all levels.

## 3 RESULTS

Figure 1 shows the instruction digram frequency distribution across the entire code sample. It can be seen that a small number of digrams occur frequently while most digrams are much less common. Over 50% of digrams occur only once and over 90% occur 10 or fewer times.

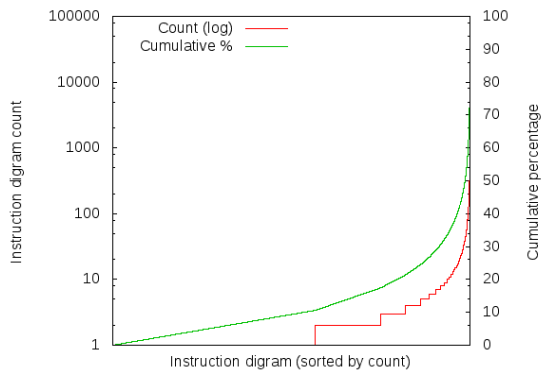

Figure 1: Instruction digram frequency distribution.

The digrams that do occur represent just 6.25% of all possible digrams. This means that 93.75% of possible digrams are not represented in the code sample at all. Figure 2 provides a visualisation of digram frequency (or rather scarcity) as a 3-d chart. The x and y axes both correspond to individual instructions and each x-y point corresponds to an ordered pair of instructions - a potential digram. The colour of each point (z-axis) represents digram frequency using a log scale to enhance the detail. It is clear from Figure 2 that many digrams are not used and most of those that are used are infrequent.

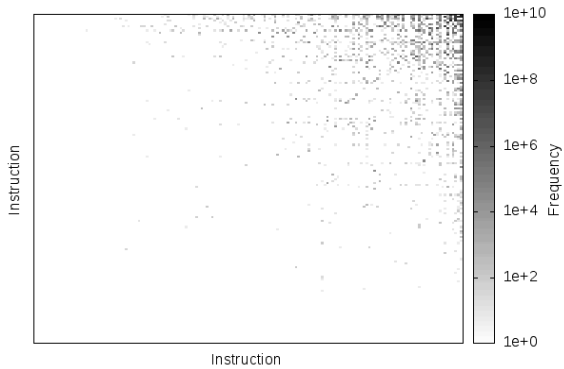

Figure 2: Instruction digram frequency visualisation.

Some of the digrams that do not occur are absent due to data type constraints. For example, an instruction that returns a string cannot be used as an input for a second instruction that expects a number. Another reason why some digrams might be missing is due to redundancy. For example, there is no point in applying one instruction to reverse a list and then another instruction to sort the list. In this scenario the effect of reverse instruction is completely undone by the sort. The same is true of pairs of instructions that have opposite effects such as conversion to uppercase and lowercase, or reverse-reverse. Functional equivalence of pairs of instructions is another possibility.

However, these explanations do not account for many of the missing potential digrams. Other possible explanations include preference, different forms of bias or the applicability of particular instructions to different kinds of problems. Regardless of the reason it is clear that the instruction digram distribution holds across a large and diverse code sample. As a result we can potentially take advantage of it to constrain search.

Instruction digrams can be applied as constraints to instruction subsets. This effectively reduces the number of ways in which instructions from the instruction subset can be combined.

Instruction subsets exist with different maximum sizes. The maximum subset size dictates an upper limit in terms of the number of distinct instructions that can appear in a solution. Note that this does not place a limit in the total number of instructions since the same instruction can, and often is, repeated within a given program unit.

Placing a limit on the number of instructions is useful given that 90% of program units contain 10 or fewer distinct instructions while only 2% of program units contain more than 20 instructions. Having smaller instruction subsets greatly reduces the overall size of the search space so if we search using smaller subsets first then we will find most solutions and do so more quickly. If this proves fruitless and we are prepared to expend the additional effort then we can try progressively larger instruction subsets until a solution is found.

Instruction digrams can reduce the size of the search space for an instruction subset since it allows us to ignore many instruction applications that we might otherwise have to consider. If we accept that we will only allow instruction applications that are defined as instruction digrams and we know that only a small proportion of possible instruction digrams exist then we can expect to create many fewer instruction applications.

We applied the instruction digrams as constraints to the instruction subsets that were produced in the previous study and measured the resulting search space size using the approach described earlier. Figure 3 shows the search space sizes for instruction subsets of size 10 - with and without instruction digrams. This chart shows data flow depth as the x-axis. This corresponds to the maximum depth of the search graph as measured in instructions from the root. The y-axis corresponds to search space size as measured in search space nodes and uses a log scale.

It can be seen that the size of the search space using all instructions (red '+' line) grows rapidly. The search space size quickly makes exhaustive search infeasible, whatever approach is used. Instruction subsets (green 'x' line) provide a significant reduction in search space size, by only using combinations of instructions that people use when writing software. Instruction subsets with instruction digrams (blue dots) can be seen to give a further significant reduction in search space size. There are around 6000 instruction subsets of size 10 so these have been jittered to improve visibility. In this case the jitter adds a small random value to the x-axis that results in the data points being distributed around the relevant line. This random value has no significance from a data perspective.

The search space sizes for instruction subsets plus digrams assume a wide range of values. This is to be expected since the subsets contain different instructions and these will correspond to different sets of digrams. Some of the instructions are common while many are rare. Rare instructions are expected to have fewer corresponding digrams. An instruction subset with few digrams will have a significantly smaller search space than one with many digrams.

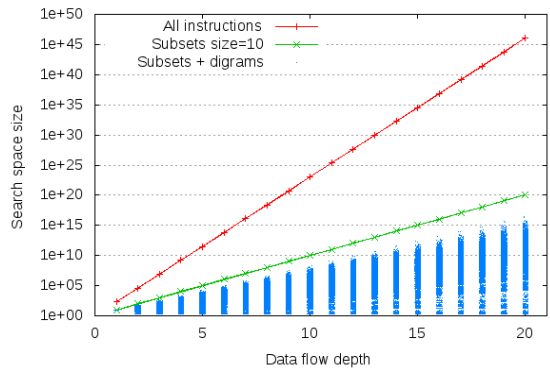

Figure 3: Search space sizes for subset size 10.

Above data flow depth of 5, the worst case (largest) search space size for digrams is at least one order of magnitude smaller than for instruction subsets alone. At greater data flow depths the size of the reduction continues to increase. We can see by tracing horizontally from the (green) subset line to the highest (blue) digram point that for the same resources instruction digrams allow deeper exploration of the search graph by between 1 and 4 levels.

Figure 4 shows a similar chart for instruction subset size 20. Here the line for all instructions is identical. Larger subsets naturally have a larger corresponding search space. As a result the instruction digram sizes are also larger. However, there is a much greater reduction in search space size. Taken together, subset sizes 10 and 20 account for 98% of the program units in the code sample.

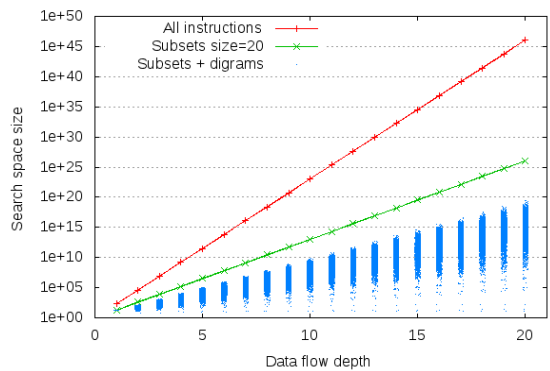

Figure 4: Search space sizes for subset size 20.

Figure 5 shows the search space sizes for subset size 50. This is included mainly to show the continuation of the trend. The reduction in search space size continues to increase. Also, it is clearer that the spread of the digram search space sizes is becoming narrower.

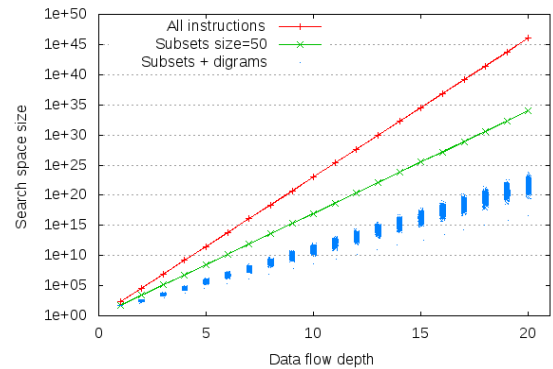

Figure 5: Search space sizes for subset size 50.

The overall distribution of search space sizes is not easily discernible from Figures 3-5. Figure 6 shows the search space sizes for instruction subsets plus digrams. The lines correspond to all of the blue dots in the previous three figures. These are plotted for different subset sizes and the individual subsets have been sorted in order of ascending search space size along the x-axis.

It is clear that there are different numbers of subsets for different subset sizes. The kink at the end of each line shows that relatively few subsets have significantly larger search space sizes than would be expected from the main sequence trend. These correspond to the highest search space sizes for data flow depth 20 in Figures 3-5 (blue dots). It should be remembered that even the largest search space sizes represent a significant reduction. The key take-away is that the majority of subsets have search space sizes that are much smaller than the maximum.

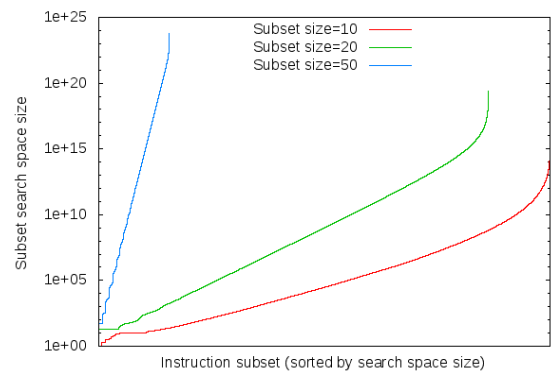

Figure 6: Reduced instruction subset search space sizes.

The reduction in search space size provided by instruction digrams is less dramatic than that for instruction subsets but it is still significant. Particularly at subset size 10, many of the individual search space sizes for subsets plus digrams are 5 or more orders of magnitude smaller that for subsets alone.

In Zoea different subsets are processed by a pool of worker nodes. The above results mean that all of the jobs would complete more quickly and perhaps half of the jobs required to cover the entire search space would also complete in a tiny fraction of the time required otherwise.

## 4  DISCUSSION

As with instruction subsets, instruction digrams can be viewed as another type of low level, tacit software development knowledge. In the same way, the identification and exploitation of instruction digrams represent a form of knowledge elicitation. Other similar forms of low level programming knowledge may also exist and if so should be investigated.

The whole approach is predicated on the assumption that there is merit in mimicking the way in which people develop software. In the absence of alternatives this is at least an expedient strategy. From another perspective, the size of the inductive programming search space has been a major and pernicious obstacle for decades. Thus any approach that significantly reduces this problem is of interest. However, there remains a risk that by excluding the consideration of many combinations of instructions we might overlook some simpler and innovative solutions to problems.

It is not known why the instruction digram frequency distribution in human code is skewed in the way that it is. This is no doubt somehow influenced by the skewed instruction distribution but why are some digrams common while many possible digrams are never used? Human patterns of code use may not be optimal and alternative digram distributions may exist that are better in some ways.

The instruction digram approach does not consider context. Larger structures such as trigrams or n-grams might be worth investigating. Alternatively, context could be applied extrinsically. Context would necessarily involve instructions that have already been hypothesised as part of the candidate solution, at lower levels in the search tree and on data flow paths that lead to the current node. A simplistic approach might utilise different instruction subsets and different sets of digrams for various domains. Indeed, it is possible that this occurs already to some extent in the case of subsets, as a result of clustering. Instructions that co-occur in particular kinds of code are likely to end up in the same subsets.

Integration of instruction subsets and instruction digrams is remarkably easy. These elements represent two different forms of knowledge:

1. What instructions to use together and
2. How to combine instructions to form programs.

Perhaps this experience can guide the integration of any additional forms of coding knowledge that are identified.

The code sample is assumed to be representative of other code. It is large, diverse and includes code relating to a wide range of domains. It is also the work of a large number of developers.

The current study only considers the direct application of instructions. It would also have been possible to include the indirect application of instructions via variables. For example the code:

```
b = int(a)
c = abs(b)
```

involves the application of two instructions (abs and int) using the intermediate variable b to effect the data flow. While useful, this approach would have added significant complexity to the parsing process. Instead, the focus on direct application is treated as a sampling exercise with the assumption that it is also representative of indirect application.

The concept of instruction digrams together with an associated frequency distribution is somewhat reminiscent of probabilistic grammars [19], which was an early line of investigation in the development of Zoea. The earlier idea was to produce a target language grammar with probabilities on the productions that reflected how people write software. The goal was to use this grammar to limit the search space size. It proved difficult to create such grammars manually so that approach was ultimately shelved. Instruction digrams represent a partial resurrection of that idea.

Zoea is transparent in the sense that it can recount and explain every step of its reasoning in producing a particular solution. Instruction digrams are compatible with this behaviour. In combination with instruction subsets it is feasible to explain the basis for the selection of each instruction in a solution.

The instruction digram approach does not currently consider the argument position in the case of instructions having multiple inputs. It is expected that this would have some marginal benefit in terms of search space size - albeit at the expense of a slightly higher operational overhead.

Digram frequency is not currently exploited in any sophisticated way. Rather it is simply used in the binary identification of digrams that do or do not exist in the code sample. Frequency could be used to threshold on some value other than 1 to exclude digrams that are rarely used. It could also be used to prioritise the application of instructions - potentially allowing many solutions to be identified sooner.

As with instruction subsets, this approach could be useful in any area where a state space of code solutions needs to be searched. It is also applicable to any kind of production system including rules and grammars – subject to the availability of a suitably annotated corpus of solutions, such as parse trees.

Most high-level languages have similar sets of instructions and it is reasonable to expect that this approach is generally applicable to imperative languages other than python. This would of course need to be verified. Similarly, it is expected that the approach would be useful in non-imperative paradigms such as logic programming.

The search space sizes presented here are pessimistic and represent upper bounds. The actual performance will only be determined with operational deployment of the approach in Zoea. Deployment of instruction subsets in Zoea has been deferred pending the results of the current study. It is anticipated that both approaches will now be deployed in the near future.

There are often many different ways of coding a functionally identical program. This means that there is a high probability that an alternative but equivalent solution will be found before any particular variant that might be anticipated. Zoea treats the first (and thus the smallest) program it finds that satisfies all of the test cases as a suitable solution.

It is worth noting that even though there are often thousands of instruction subsets of a given size, most solutions are found in the first 10 subsets. This is due to the significant overlap between subsets. The majority of instruction subsets exist to account for the long tail of increasingly rare combinations of instructions. Programs that utilise particularly uncommon combinations of instructions can expect to require search involving many or most of the subsets. In these cases the benefits of instruction digrams are expected to be highly significant. Long tail subsets tend to have much smaller search spaces due to the fact that uncommon instructions have many fewer digrams.

As far as the authors can tell the key ideas behind this study are novel. Given the relative simplicity of the approach it is surprising that similar work has not been done before. Based on this study and the previous one it would seem there is still much to be learned from a deeper analysis of code.

We believe that the statistical analysis of instruction digrams in open source code to be ethical. Digrams cannot be considered to represent anybody's intellectual property, as many of them are not unique to any particular program. Also, the process does not extract, store or use any executable fragments of code. Other than in trivial cases, it would be impossible to infer anything from the digrams about the algorithms present in the code sample. In any event we are only interested in the digram frequency distribution across the complete code sample.

## 5 CONCLUSIONS

We have described the use of instruction digrams as a heuristic for search in inductive programming. We analysed the instruction digram frequency distribution within a large sample of open source code. It was noted that many possible instruction digrams are not represented in the code sample. The distribution of instruction digrams that were present is highly skewed with only a small number being very common. We modelled the search space size for instruction subsets of different sizes, constrained by instruction digrams. It was found that instruction digrams yield a further reduction in search space size of up to several orders of magnitude. This will allow a search graph to be explored more deeply with the same time and resources, and means that larger programs can be produced as a result. We have discussed the results and identified some opportunities for further work.

## ACKNOWLEDGEMENTS

This work was supported entirely by Zoea Ltd. Zoea is a trademark of Zoea Ltd. Any other trademarks are the property of their respective owners. Copyright © Zoea Ltd. 2023. All rights reserved.